\documentstyle[epsf,epsfig,rotate]{mn}

\newcommand{\ls}{\raisebox{-.8ex}{$\buildrel{\textstyle<}\over\sim$}}

\newcommand{\anrev}{{\it ARA\&A, }}
\newcommand{\apj}{{\it ApJ, }}

\newcommand{\icar}{{\it Icarus, }}
\newcommand{\mnr}{{\it MNRAS, }}
\newcommand{\nat}{{\it Nat, }}

\title{Dynamical Relaxation and Massive Extrasolar Planets}

\author[J.C.B. Papaloizou \& C. Terquem] {John C. B. Papaloizou \\
       Astronomy Unit, Queen Mary \& Westfield College, Mile End Rd,
       London E1~4NS, UK -- jcbp@maths.qmw.ac.uk \and Caroline Terquem
       \\ Institut d'Astrophysique de Paris, 98~bis Boulevard Arago,
       75014 Paris, France -- terquem@iap.fr \\ Universit\'e Denis
       Diderot--Paris VII, 2 Place Jussieu, 75005 Paris, France}

\date{Submitted to MNRAS on December 18, 2000}

\pubyear{}

\begin{document}

\maketitle

\begin{abstract}
Following the  suggestion  of Black(1997) that some massive extrasolar planets
may  be associated with the tail of the distribution of stellar companions, 
we  investigate  a scenario in which
 $5 \le N \le 100$ planetary mass  objects  are assumed
to form  rapidly through a fragmentation process
occuring in a disc  or protostellar envelope on a scale of 100~au.
 These are assumed to have formed rapidly enough through gravitational
instability or fragmentation that their orbits can undergo dynamical
relaxation on a timescale $\sim 100$ orbits.

 Under a wide range of initial conditions and assumptions the
relaxation process ends with either (i) one potential 'hot Jupiter'
plus up to two 'external' companions, i.e.  planets orbiting near the
outer edge of the initial distribution; (ii) one or two 'external'
planets or even none at all; (iii) one planet on an orbit with a
semi--major axis 10 to a 100 times smaller than the outer boundary
radius of the inital distribution together with an 'external'
companion.  Most of the other objects are ejected and could contribute
to a population of free floating planets.  Apart from the potential
'hot Jupiters' , all the bound objects are on highly eccentric orbits.
We found that, apart from the close orbiters, the probability of
ending up with a planet orbiting at a given distance from the central
star increases with the distance.  This is because of the tendency of
the relaxation process to lead to collisions with the central star.
     
We discuss the application of these results to some of the more
massive extrasolar planets.

\end{abstract}


\section{Introduction} \label{intro}


The recent discovery of extrasolar giant planets orbiting around
nearby solar--type stars (Marcy \& Butler 1998, 2000; Mayor \& Queloz
1995) has stimulated renewed interest in the theory of planet formation.
The objects observed so far have masses, $M_p$, that are
characteristic of giant planets, i.e. $0.4 \; {\rm M}_J \; \ls \; M_p
\ \ls \; 11 \; {\rm M}_J$, M$_J$ denoting a Jupiter mass.  The orbital
semi-major axes are in the range $0.04 \; {\rm au} \; \ls \; a \; \ls
\; 2.5 \; {\rm au}$, and orbital eccentricities in the range $ 0.0 \;
\ls \; e \; \ls \; 0.67$ (Marcy \& Butler 2000).

It is a challenge to formation theories to explain the observed masses
and orbital element distributions.  There are two main theories of
giant planet formation (see Papaloizou, Terquem  \& Nelson 1999 and references
therein).  One is the core instability scenario. In this, a solid core
of several earth masses is built up in the protostellar disc, at which
point it is able to begin to accrete gas and evolve to become a giant
planet. Once massive enough it is able to open a gap in the disc and
undergo orbital migration through disc protoplanet interactions
(e.g. Lin \& Papaloizou 1993).  It has been suggested that the high
orbital eccentricities for extrasolar planets might be explained by
disc-protoplanet interactions (Artymowicz 1992).

Recent simulations of protoplanets in the observed mass range (Kley
1999, Bryden et al. 1999, Lubow, Seibert \& Artymowicz 1999)
interacting with a disc with parameters thought to be typical of
protoplanetary discs, but constrained to be in circular orbit,
indicate gap formation and upper mass limit consistent with the
observations.  However, simulations by Nelson et al. (2000) that
relaxed the assumption of fixed circular orbits found inward migration
and that the disc-protoplanet interaction leads to strong eccentricity
damping.  Thus the observed eccentricities of  apparently isolated extrasolar planets
are
so far unexplained by this scenario.

The other possible formation mechanism is through fragmentation or
gravitationnal instability in a protostellar disc (e.g. Cameron 1978,
Boss 2000). This may occur early in the life of a protostellar disc
surrounding a class~0 protostar on a dynamical timescale.  Such discs
have been observed (see, e.g., Pudritz et al. 1996) and the
characteristic size is about 100~au.  It is unlikely that such a
process would operate at distances smaller than about 50~au from the
central star as, in the optically thick parts of the disc, non
axisymmetric density waves redistibute mass  and angular momentum before
fragmentation can proceed ( eg. Papaloizou \& Savonije 1991, Laughlin \& Bodenheimer 1994).
Fragmentation  is more likely when cooling is efficient, as may
occur in the optically thin parts of the disc, beyond about 50~au
(Papaloizou et al~1999).   However, the detailed conditions
required for it to occur are unclear and may require constraining
influences from the external environment ( Pickett et al 2000).
Note that fragmentation may also occur
before a disc is completely  formed, during the initial collapse of the
protostellar envelope.  Such opacity limited fragmentation has  been
 estimated to produce objects with  a lower mass limit of  7 Jupiter masses
(Masunaga~\& Inutsuka 1999), but there is no definitive argument to
rule out somewhat smaller masses (Bodenheimer et al. 2000). 
It is
possible that both a disc and fragments may form simultaneously out of
the envelope, the relative importance of the two processes depending
for instance on the angular momentum content of the envelope, on the
strength of any magnetic field ( so far neglected  in disc fragmentation
calculations) and possibly on the initial clumpiness.
Note that large scale  observations of class~0 envelopes so far do not rule out the
presence of clumps with masses smaller than about 10 Jupiter masses
(Motte \& Andr\'e 2001).

It is the purpose of this paper to investigate the evolution under
gravitational interactions of a distribution of $N$  massive planets which
we assume to have been formed through a fragmentation process rapidly
enough that their orbits can undergo  subsequent dynamical relaxation
on a time scale of hundreds of orbits. In common with related work
on orbital evolution ocuring after assummed formation in a disc
(eg. Weidenschilling \& Mazari  1996, Rasio \& Ford  1996,  Lin \& Ida 1997)
we shall neglect the effects of any remnant disc gas so that apart from tidal
interactions with the central star there are only gravitational interactions.
Thus this work  complements studies of the initial fragmentation process
in a gaseous medium. 

It turns out that the resulting evolution leads to similar
end states   independently  of whether 
the initial  configuration  is assumed to be in the form  of a  spherical  shell
or a disk--like
structure.  

The motivations for this work are  firstly the suggestion by Black (1997)
and Stepinski~\& Black (2000) that massive extrasolar planets on
highly eccentric orbits could actually be the low--mass tail of the
low--mass companion distribution to solar--like stars produced by fragmentation
processes. Here we wish to investigate to what extent planets with orbital
elements similar to those observed can be produced and in particular
whether 'hot Jupiters' orbiting close to the star can be formed. Secondly  we consider the
 recently detected population of free-floating
planets and  its relationship to that of planets orbiting solar--type stars
( Lucas \&
 Roche 2000, Zapatero Osorio et al. 2000). It is of interest to know
to what extent free-floating planets could be produced as a result of
ejection  from the neighbourhood of a star.

We have considered the orbital evolution of $N$ bodies with masses in
the giant planet range, which are assumed to be formed rapidly, using
up the gas in a protostellar disc or envelope around a solar mass
star, so that they can undergo subsequent dynamical relaxation on a
timescale $\sim 100$ orbits.  We have performed calculations with $5
\le N \le 100.$ 

In all the runs we performed, most of the planets where ejected from
the system and at most 3 planets remained bound to the central star.
We found that close encounters with the central star occured
for about 10\% of the planets for all values of $N$ considered.  
Such close encounters early in the evolution tended to result in 
 collisions.  These  tended to be avoided at later times
 so that tidal interaction   might then result in orbital
circularization at fixed pericentre distance 
leading  to the formation of a very closely orbiting giant planet.

Typically the runs ended up with either (i) one potential 'hot Jupiter' plus up to
two 'external' companions, i.e.  planets orbiting near the  outer edge of the
initial distribution; (ii) one or two  'external' planets  or even none at
all; (iii) one planet on an orbit with a semi--major axis 10 to a 100
times smaller than  the outer boundary radius of the
inital distribution together with an 'external' companion.
  Apart from the potential  'hot
Jupiters' , all these objects are  on highly eccentric orbits.  We found
that, apart from  the close orbiters, the probability of ending up with a
planet orbiting at a given distance from the central star increases
with the distance.

The objects that become unbound may contribute to a population of
freely floating planets ( Lucas \&  Roche 2000, Zapatero Osorio et al.
2000) which could be several times larger than that of giant planets
found close to the central star.
     
Thus the dynamical relaxation process considered here may operate in
some cases to produce giant planets with high orbital eccentricity at
several astronomical units from their central star as well as closely
orbiting planets.  In all cases a population of loosely bound planets
is also expected.
 
The plan of this paper is as follows.  In section \ref{mod} we
describe the model and basic equations used.  In section \ref{ini-bc}
the initial conditions are formulated and in section \ref{relax2} the
physics of the relaxation process is discussed.  In section
\ref{results} we present our numerical results and in section
\ref{disc} we summarize and discuss them.

\section{The model and basic equations} \label{mod}

We consider a system consisting of $N$ planets and a primary star
moving under their gravitational attraction.

\noindent As we are interested in possible close approaches to, or
collisions with, the central star we adopt a spherical polar
coordinate system $ (r, \theta, \varphi)$ with origin at the
stellar centre.  The planets and central star are treated as point
masses.  However, to take into account possible losses of orbital
energy and angular momentum to the stellar material, a simple model
for taking into account the tidal interaction between the star and a
closely approaching planet is also included.

The equations of motion are \begin{equation} {d^2 {\bf r}_i\over dt^2}
= -{GM_*{\bf r}_i\over |{\bf r}_i|^3} -\sum_1^N {GM_j{\bf r}_{ij}
\over |{\bf r}_{ij}|^3} -{\bf a} +{\bf F}_{Ti}. \label{emot}
\end{equation} Here $M_*, M_i, {\bf r}_i$ and ${\bf r}_{ij}$ denote
the mass of the central star, the mass of planet $i,$ the position
vector of planet, $i,$ and ${\bf r}_i-{\bf r}_j$ respectively.  The
acceleration of the coordinate system based on the central star
(indirect term) is \begin{equation} {\bf a}= \sum_1^N {GM_j{\bf r}_{j}
\over |{\bf r}_{j}|^3}
\label{indt}\end{equation} and that due to tidal interaction with planet
$i$ is ${\bf F}_{Ti}.$

In the situation envisaged here, tidal interactions occur when a
planet has a close encounter with the star.  When this occurs, the
planet approaches from large distances on an almost parabolic
orbit. The time between a subsequent encounters will then be long
compared to that for the tidal interaction itself.  Accordingly we
approximate the process as a sequence of independent energy and
angular momentum transfers that occur at each periastron passage.  We
utilize the results of Press \& Teukolsky (1977) who calculated these
transfers in the small perturbation limit for a non rotating star
modelled as a polytrope and a perturber on a parabolic orbit. We shall
neglect the effects of tides acting on the planet itself. Accordingly
our model is simplified. However, it does enable us to include tidal
effects and demonstrate how they start to lead to orbital
circularization and gravitational decoupling of an inner planet from
the others as it moves onto a close orbit. However, it is only
applicable while the planet orbit has high eccentricity.

We adopt a form for $ {\bf F}_{Ti}$ that is able
to  approximately give the correct angular momentum
and energy exchanges with the star on a close approach
but which is  negligible at
larger distances from the central star.

\begin{equation} 
{\bf F}_{Ti} = -{GM_{i}R_*^5R_{pi}^3T_1\over C_1 |j_i| |{\bf
 r}_i|^{11}} (|{\bf r_i}|^2{\bf v}_i -{\bf v}_i\cdot{\bf r}_i{\bf
 r}_i) \label{ftidal}.\end{equation}

Here $j_i$ is the specific angular momentum of planet $i,$
$R_*$ is the stellar radius, $R_{pi} =j_i^2/(2GM_*)$ is
the distance of closest approach of planet $i$ assuming
a parabolic orbit, $C1=2\sqrt{\pi}/3,$ and $T_1= 0.6/(1+(R_{pi}/R_*)^3).$

The equations of motion are here integrated using the
Bulirsch-Stoer method ( eg. Press et al 1993).

Using (\ref{ftidal})   we can derive the energy lost to the star
during a close encounter  of planet $i$, assumed on a parabolic orbit as
\begin{equation} 
\Delta E = - \int^{\infty}_{-\infty}M_{i}{\bf F}_{Ti}\cdot {\bf v}_idt
\end{equation} where, because of the rapid convergence of the
integral, the limits are extended to $\pm \infty.$ This gives
\begin{equation} 
\Delta E = { GM_{i}^2 R_*^5T_1\over R_{pi}^{6}},
\label{DE} \end{equation} which gives values coinciding approximately
with values given by Press \& Teukolsky(1977).

We comment that acording to (\ref{DE}) a star grazing encounter
on a parabolic orbit results in a final semi-major axis
$a \sim 1.7R_*M_*/M_i.$ For a central solar mass with $R_*=10^{11} cm.$
and $M_i \sim 1M_J,$ $a \sim 10$~au. Thus bound orbits
with $a\sim 10$~au are significantly modified if they have
a close encounter with the central star. Note too that the energy
exchange rates are small for the planetary mass objects
considered here  giving some justification for the
linear approximation used to calculate them.

\section{Initial  Conditions}
\label{ini-bc}

The simulations performed here were begun by placing $N$ planets in
some specified volume in a random location chosen to give a
Monte-Carlo realization of a prescribed density distribution.  We
considered both the case of a uniform density spherical shell with
$R_{min} \leq r \leq R_{max}$ and that of a thick annulus with
$R_{min} \leq r \leq R_{max}$ occupying $\cos^{-1}(0.1) \leq \theta
\leq \cos^{-1}(-0.1) $ with a density $\propto r^{-2}$, where $r$ is
the spherical radius and $\theta$ the co--latitude.  For the
calculations presented here, $R_{min}= 0.1 R_{max}$
and $0.5~R_{max}$ for the annulus and the spherical shell,
respectively.  The planets were then given the local circular velocity
in the azimuthal direction.  Note that because of the spatial
placements the initial orbits are not coplanar.  In some cases the
planets were taken to have equal mass $M_p$ while in others each
planet was allocated a mass $qM_p$ where $q$ was a random number
between zero and one.  In the latter case the mass distribution of the
protoplanets does not correspond to the specified density distribution
indicating some redistribution of mass among the embryos.

In general the various initial conditions we have considered lead to
the same qualitative evolution.

However, it must be emphasized that the systems are chaotic with the
consequence that very small detailed changes to the initial conditions
or the integration procedure will in general lead to very different
results in detail. To deal with this issue one can adopt the notion of
shadow orbits (Quinlan \& Tremaine 1992).  According to this, one can
expect that although inevitable small errors lead to a significant
deviation of the numerical solution from the actual one, there is
another solution of the real system obtained with slightly different
initial conditions that remains close to the numerical one.  Thus we
should be able to identify qualitative trends in the evolution of an
ensemble of systems of the type we consider and we restrict ourselves
to that.

\subsection{Scale Invariance}

We comment that the equations we solve, incorporating tidal effects,
have a radial scale invariance.  Thus all radii may be mutiplied by
some factor $f$ while the timescales are multiplied by $f^{3/2}.$ The
size of the central star has to be scaled by the factor $f$ also.
Here we shall take the unit of length to be $R_{max}$.  Then the
stellar radius is specified through $R_*/R_{max}.$ The time unit is
the period of a circular orbit at $R_{max},$ $P_0 = 2\pi
R_{max}^{3/2}/\sqrt{GM_*}.$ The interactions amongst the planets lead
to some escaping the system.  Objects with positive energy which had
reached a distance $\beta R_{max}$ were considered to be escapers.  We
have considered $\beta=20$ and $\beta=100$ which both lead to the
same qualitative picture.

\section{Evolution relaxation and star grazers} 
\label{relax2}

Even the systems which have a small number of bodies are found to
interact strongly and to undergo relaxation like a stellar system
(Binney \& Tremanine 1987).  For such a system the realaxation time is
\begin{equation} t_R ={0.34 v^3\over 3\sqrt{3} G^2 M_P\rho
\ln(\Lambda)}\label{relax}\end{equation} Here the root mean square
velocity is $v,$ $\rho$ is the mass density of interacting bodies
assumed for simplicity to have equal mass $M_p$, and $\Lambda
=M_*/M_p.$

Using $v^2 = GM_*/R,$ $NM_p= 4\pi R^3 \rho/3 $ and adopting
the orbital period $P = 2\pi R/v$ this becomes

\begin{equation} 
t_R={0.044M_*^2 P\over M_p^2 N \ln (M_*/M_p)}.\label{relax1}
\end{equation}

Thus for $M_p/M_* =5.0\times 10^{-3}$ and  $N=5$ the relaxation time
is about $100$ orbits. For systems with $R$ in the $100$~au to $1000$~au
range this time is around $10^{5-6}$~y which is within 
the estimated lifetimes of protostellar discs.

The evolution we obtain is similar to that undergone by spherical star
clusters (Binney \& Tremaine 1987).  The relaxation, due to binary
encounters, causes some objects to attain escape velocity while others
become more bound.  Eventually all either escape or end up in extended
orbits, except for one body which takes up all the binding energy.
This is a generic result provided that close encounters with the star
can be avoided.  However, for the parameter range of interest such
encounters are likely.  The situation resembles that of accretion of
stars from a spherical star cluster by a black hole (eg. Frank \& Rees
1976).  At a location with radius $R,$ the fractional volume of phase
space containing orbits that would collide with the star if
unperturbed is $\sim R_*/R $ ( this being the ratio of the square of
the angular momentum below which an impact is expected to the square
of the mean angular momentum).  If an object cannot diffuse out of
this volume before impact, the mean time before diffusion into the
effective volume produces an impact, for a particular object, is $\sim
t_R.$ However, the time to diffuse out of the effective volume is
$\sim (R_* t_R)/R.$ If this time is less than the crossing time $t_c
\sim R/v,$ the expected mean time to impact is $ t_{enc} = (R
t_c)/R_*.$ This time is just the crossing time divided by the
probability of being in the effective volume of phase space which is
regularly sampled because of the effective diffusion (see Frank \&
Rees 1976).

In our calculations  the innermost most tightly bound object
undergoes relaxation or phase space diffusion
 which can lead to close encounters.  For $R=25$~au, $t_c\sim 20$~y,
and $R_* \sim  1.5\times 10^{11}$~cm, $t_{enc} \sim  5\times 10^4$~y. 
Thus we can expect close encounters to occur
within the general relaxation process.
In some cases these can lead to a strong tidal interaction
which can circularize the orbit and potentially
lead to the formation of a 'hot Jupiter'.

\section{ Numerical results} 
\label{results} 

We here describe a sample of our results which illustrate the
characteristic behaviour exhibited by the systems we consider.  The
calculations presented are listed in table~\ref{tab1}.  We note that
if $R_{max}$ =100~au, $R_* / R_{max}=9.396\times 10^{-5}$ or
$1.337\times10^{-4}$ corresponds to $R_*=2$ or 3~R$_{\sun}$,
respectively, which is the radius of a protostar with a mass around
1~M$_{\sun}$ in the early stages.

\begin{table}
\begin{center}
\begin{tabular}{lllll} \hline \hline
$Run$ & $N$ & $R_*/R_{max}$         & $In$ & $M_p/M_*$  \\ \hline \\
  1   & 5   & $0.0$                 & I2   & $5\times10^{-3}$     \\
  2   & 10  & $9.396\times 10^{-5}$ & I1   & $5\times10^{-3}$     \\ 
  3   & 10  & $1.337\times 10^{-4}$ & I2   & $5\times10^{-3}$     \\ 
  4   & 8   & $1.337\times 10^{-4}$ & I2   & $5\times10^{-3}$ (R) \\
  5   & 40  & $9.396\times 10^{-5}$ & I1   & $5\times10^{-3}$     \\
  6   & 100 & $9.396\times 10^{-5}$ & I1   & $1\times10^{-2}$ (R) \\
\hline
\hline
\end{tabular}
\end{center}
\caption{ \label{tab1} This table lists the parameters for each
simulation: the number of planets $N$, the ratio of stellar radius to
$R_{max}$ and the initial set up $In,$ where $n=1,2$ denotes spherical
shell and thick annulus, respectively. The last column contains
$M_p/M_*,$ with (R) denoting that the masses were allocated uniformly
at random in the interval $(0;M_p).$}


\end{table}       

\begin{figure}
\centerline{
\epsfig{file=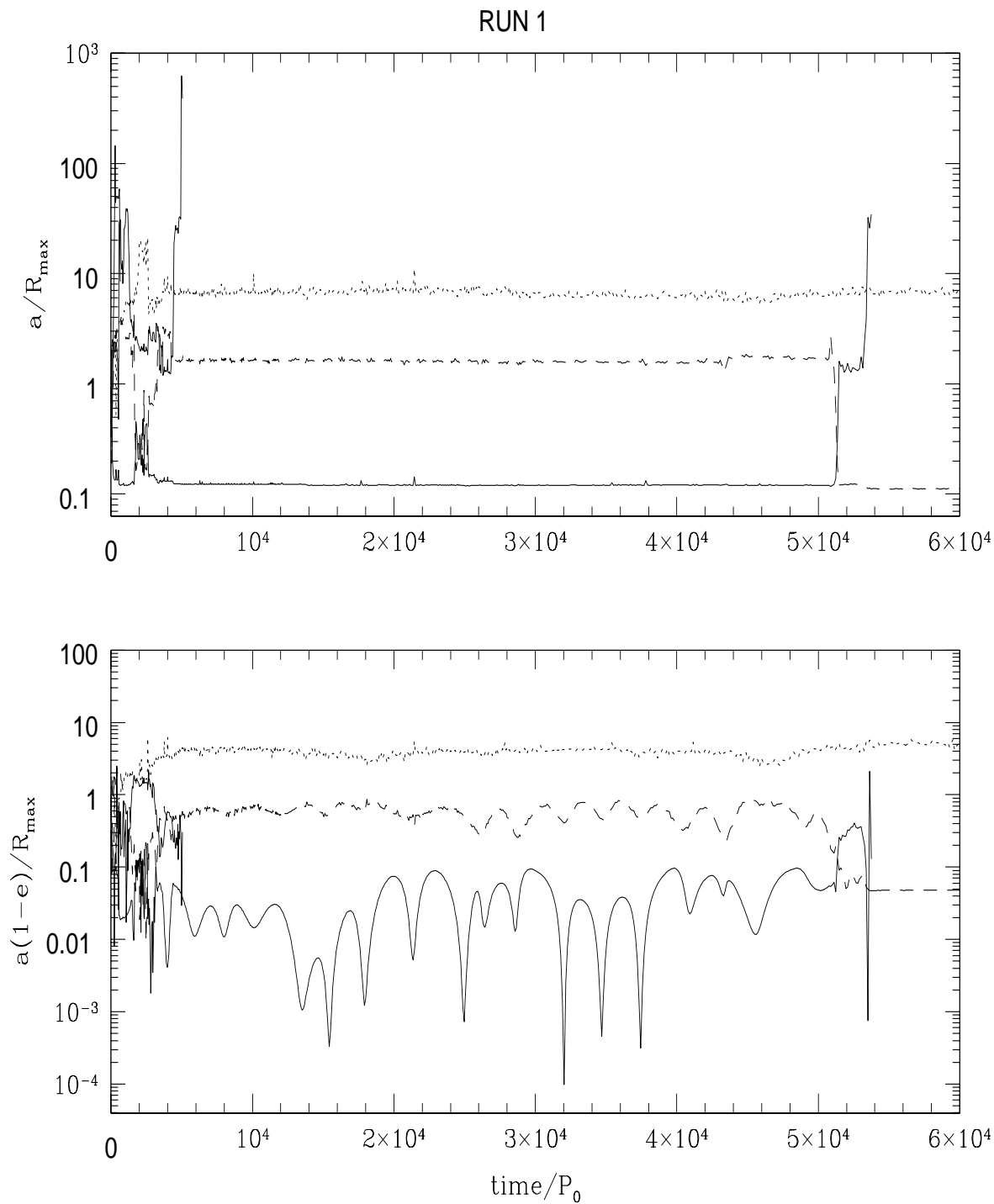,height=20.cm,width=16.cm} }
\caption[]{ This figure shows the evolution of the semi-major axes
({\em upper plot}) and pericentre distances ({\em lower plot}) of the
5 planets in the system versus time (measured in units of $P_0$) for
run 1 in table~\ref{tab1}.  The lines correspond to the different
planets each having a mass $5\times 10^{-3} M_*.$ In this and other
similar figures, a line terminates just prior to the escape of a
planet.  In this case $R_*=0$ so that there was no tidal interaction
with the central star.  Note that the time resolution in this and
similar figures is not fine enough for all close approaches to the star
to be fully  represented.  }
\label{fig1}
\end{figure}

Run~1 corresponds to a system of $N=5$ planets each having a mass
$5\times 10^{-3} M_*.$ Figure \ref{fig1} shows plots of $a/R_{max}$ in
logarithmic scale, where $a$ denotes the semi-major axis, for each
planet in the system versus time (measured in units of $P_0$).  Each
line corresponds to a different planet.  During the run 3 planets
escape, a line terminating just prior to an escape.  For this case
$R_*=0$ so that there was no tidal interaction with the central star.
The initial relaxation time for this and other similar cases is on the
order of $100P_0$, in agreement with equation~(\ref{relax1}).  The
evolution of $a(1-e)/R_{max},$ $a(1-e)$ being the pericentre distance
and $e$ being the eccentricity, is also shown in Figure~\ref{fig1}.
We see that approaches to within $ \sim 10^{-2}R_{max}$ of the central
star occur for the innermost object on a timescale of the same order
of magnitude as the relaxation time.  This is a common feature of the
simulations presented here.  Note that for $R_{max} = 100$~au, the
closest approach is to within $\sim 3\times 10^{11}$~cm, which is
comparable to a solar radius.  After about $6,000 P_0$, the main
relaxation is over with nearly all the binding energy being contained
within one object with $a\sim 0.1 R_{max}$ and $e \sim 0.9.$ However,
After about $5\times 10^4 P_0$, the two innermost planets have a close
approach which results in their position being exchanged and one of
them being ejected.  The end result of this run is then two planets on
well separated orbits.  The innermost planet has $a\sim 0.1 R_{max}$
and $e \sim 0.5$

\begin{figure}
\centerline{
\epsfig{file=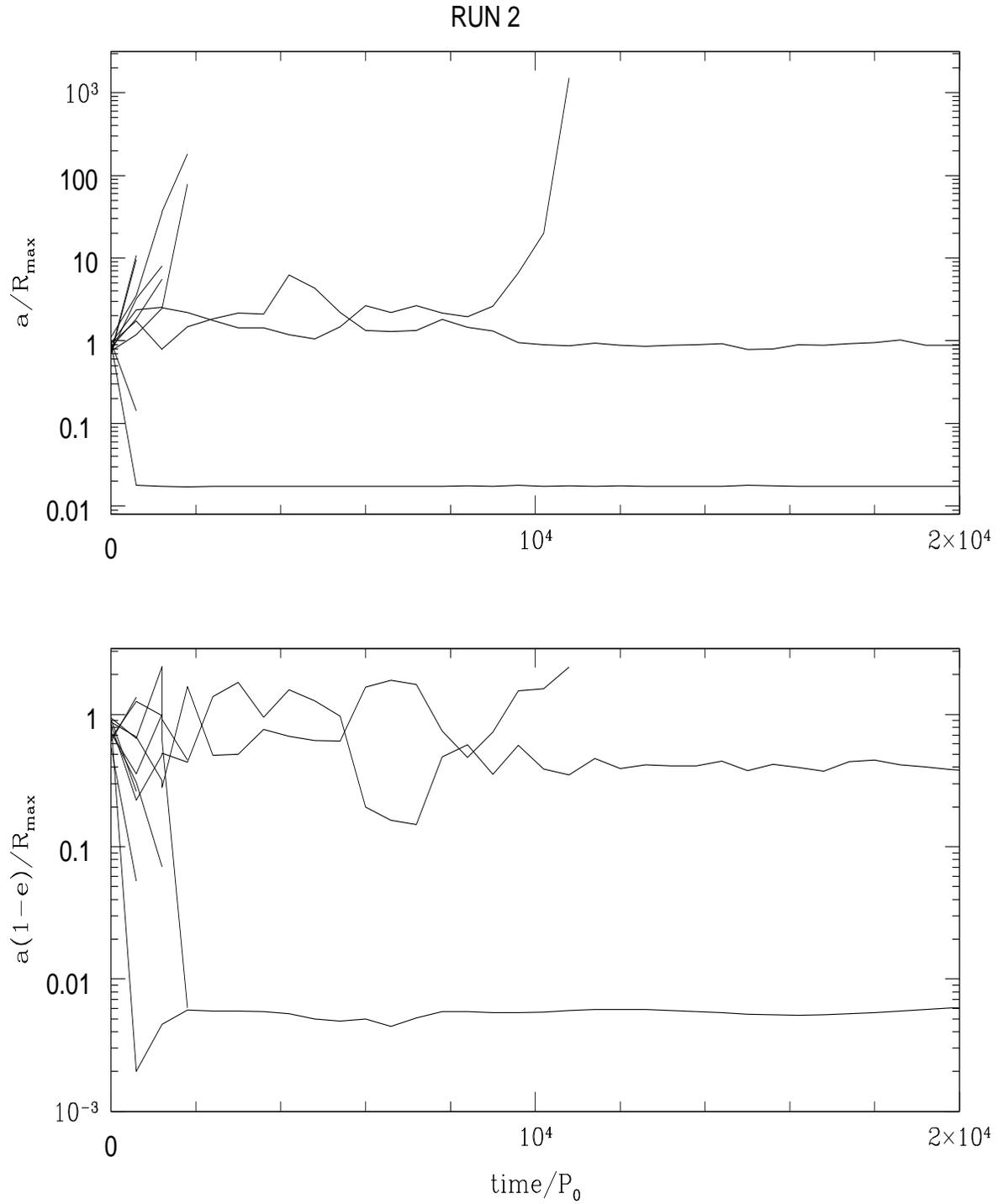,height=20.cm,width=16.cm} }
\caption[]{ Same as Fig.~\ref{fig1} but for run~2 in
table~\ref{tab1}. For this run $N=10$ and $R_*/R_{max}= 9.396\times
10^{-5}.$}
\label{fig2}
\end{figure}

A plot of the evolution of the semi--major axes and pericenter
distances versus time (measured in units of $P_0$) for run~2 in
table~\ref{tab1} is given in Figure~\ref{fig2}.  For this run the
initial number of planets is $N=10$ and $R_*/R_{max}= 9.396\times
10^{-5}$ so that tidal effects operate.  After about $10^4 P_0$, there
are only two planets still bound to the star.  The others have
been ejected.   The innermost planet which
is left has $e=0.66$ and $a \sim 0.02 R_{max}$.  If $R_{max}=100$~au,
this corresponds to $a \sim 2$~au.

\begin{figure}
\centerline{
\epsfig{file=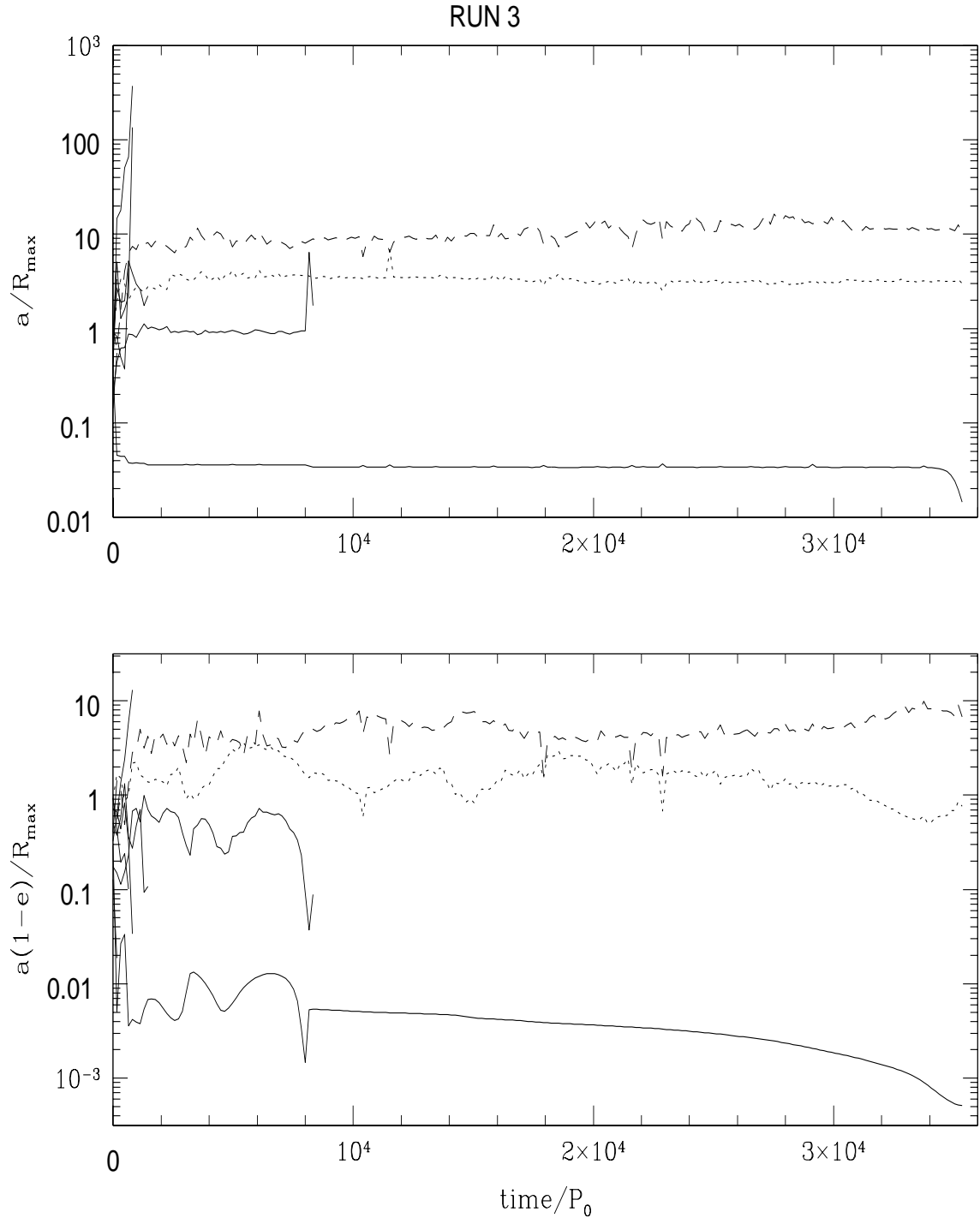,height=20.cm,width=16.cm} }
\caption[]{ Same as Fig.~\ref{fig1} but for run~3 in table~\ref{tab1}.
For this run $N=10$ and $R_*/R_{max}= 1.337\times 10^{-4}.$ The plot
terminates just before the innermost planet has a close encounter with
the star.}
\label{fig3}
\end{figure}

\begin{figure}
\centerline{
\epsfig{file=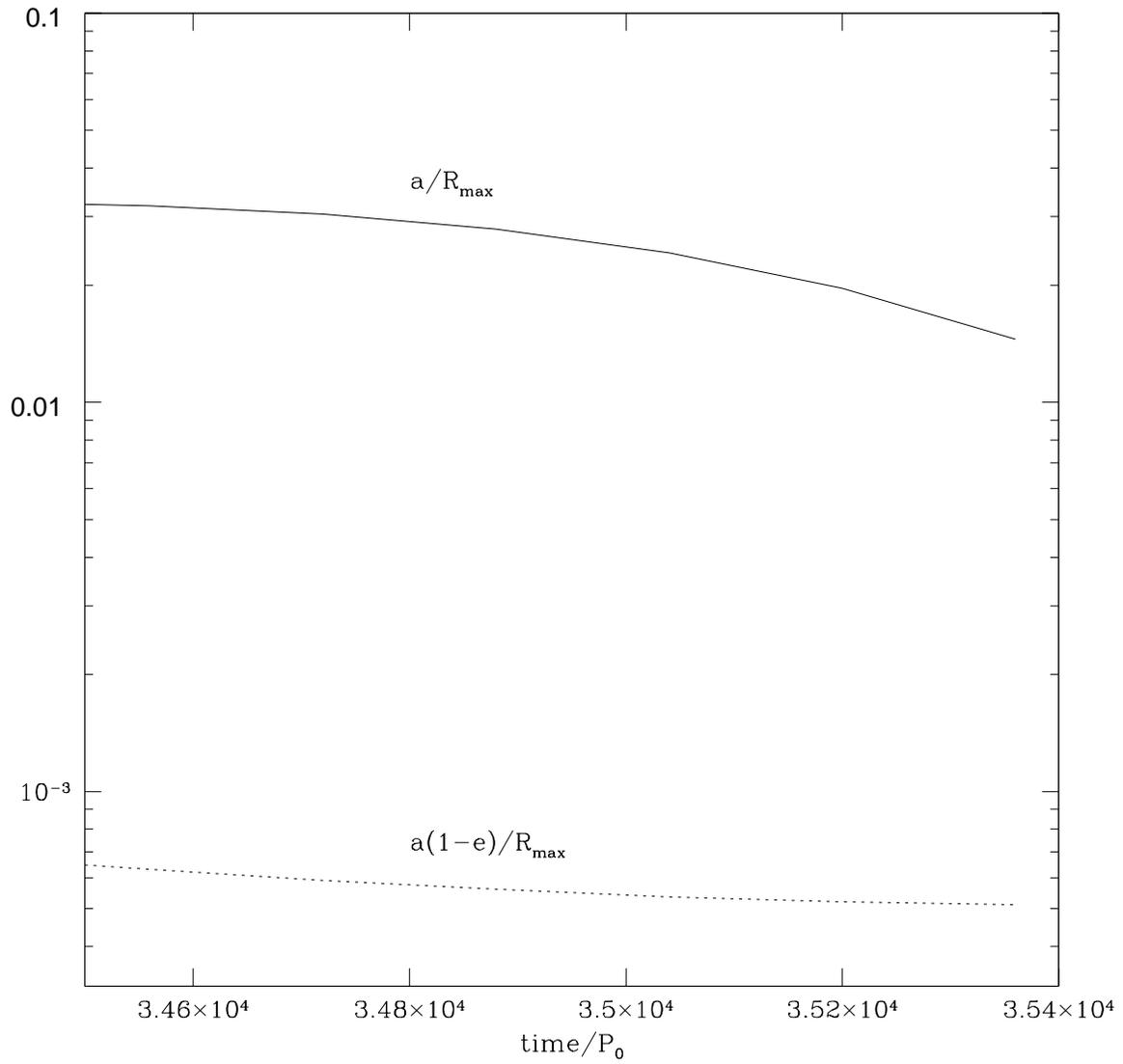,height=16.cm,width=16.cm} }
\caption[]{ This figure shows the evolution of the semi-major axis
({\em solid line}) and the pericenter distance ({\em dotted line}) of
the innermost planet in run~3 vs. time (measured in units of $P_0$)
for a short interval after this planet enters a tidal interaction
phase.  This is a zoom on the curves displayed in figure \ref{fig3}.}
\label{fig3b}
\end{figure}

Figure~\ref{fig3} shows the evolution of the semi--major axes and
pericenter distances versus time (measured in units of $P_0$) for
run~3 in table~\ref{tab1}.  For this run $N=10$ and $R_*/R_{max}=
1.337\times 10^{-4}.$ The plot terminates at about $3.4\times10^4
P_0$, just before the innermost planet has a close encounter with the
star.  The evolution toward the end of the run is zoomed on in
Figure~\ref{fig3b}.  We see that the semi--major axis decreases
significantly, down to about $10^{-2}R_{max}$, whereas the pericenter
distance varies much less.  This indicates a tidal interaction with
the star.  The run if continued much further becomes innaccurate
because of our crude treatment of tides, as mentionned in
section~\ref{mod}.  We expect at that point the orbit to circularize
at fixed pericentre distance. Evolution of this type which is also
quite common in our simulations could potentially lead to a 'hot
Jovian mass planet'.

\begin{figure}
\centerline{
\epsfig{file=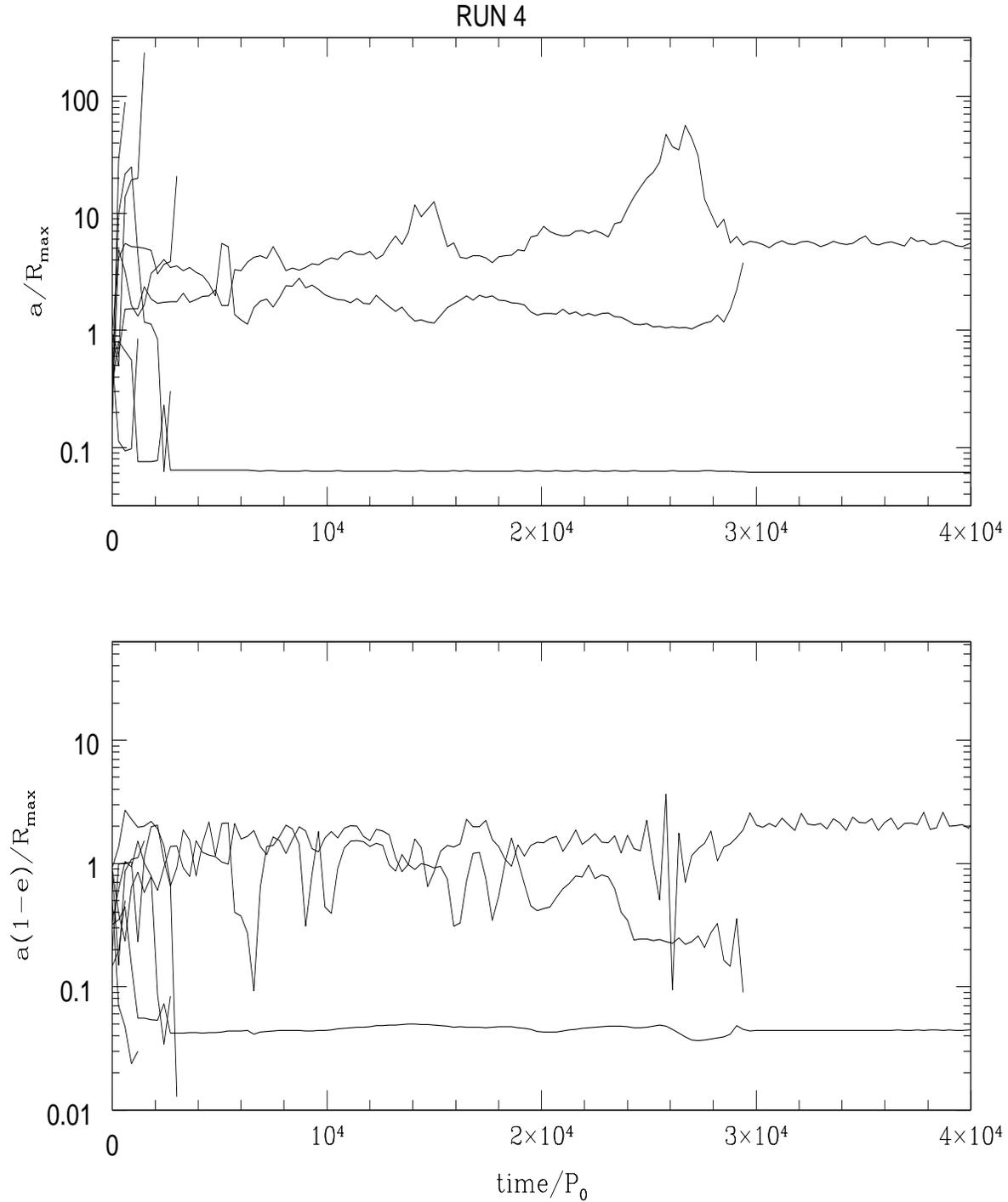,height=20.cm,width=16.cm} }
\caption[]{Same as Fig.~\ref{fig1} but for run~4 in table~\ref{tab1}.
For this run $N=8$, $R_*/R_{max}= 1.337\times 10^{-4}$ and the masses
were selected uniformly at random in the interval $(0;5\times 10^{-3}
M_*).$ At the end of the run only $2$ planets remain bound to the
central star.}
\label{fig4}
\end{figure}

Figure~\ref{fig4} shows the evolution of the semi--major axes and
pericenter distances versus time (measured in units of $P_0$) for
run~4 in table~\ref{tab1}.  In this case $N=8$,$R_*/R_{max}=
1.337\times 10^{-4}$ and the masses were selected uniformly at random
in the interval $(0;5\times 10^{-3} M_*).$ Here again, after about
$3\times10^4 P_0$ there are only two planets still bound to the star.
The innermost planet has $e=0.25$ and $a \sim 0.04 R_{max}$, which
would correspond to 4~au in a 100~au annulus.

\begin{figure}
\centerline{
\epsfig{file=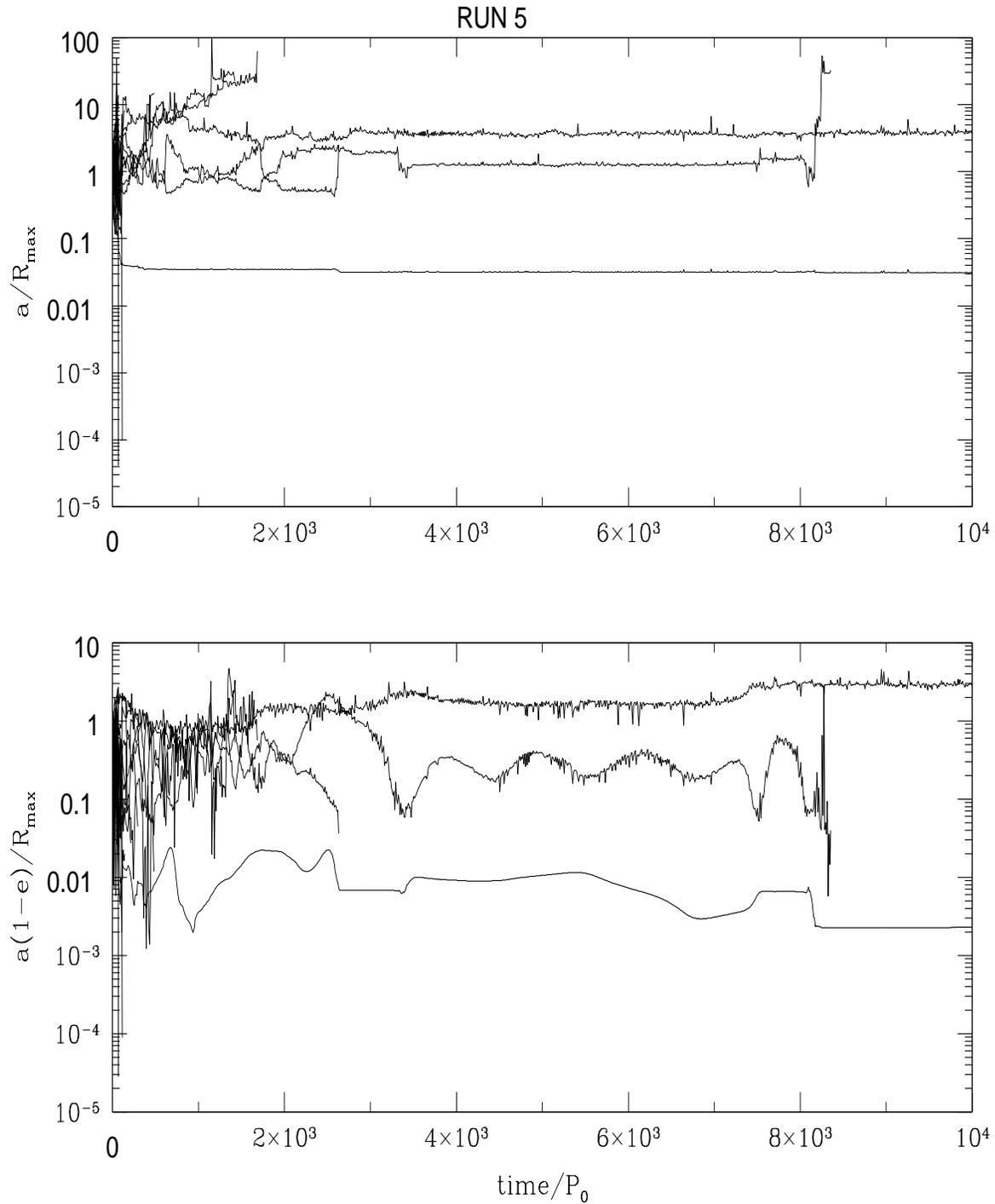,height=20.cm,width=16.cm} }
\caption[]{Same as Fig.~\ref{fig1} but for run~5 in table~\ref{tab1}.
For this run $N=40$ and $R_*/R_{max}= 9.396\times 10^{-5}$.  At the
end of the run only $2$ planets remain bound to the central star.}
\label{fig5}
\end{figure}

We now consider runs with larger $N.$ Figure~\ref{fig5} shows the
evolution of semi-major axes and pericenter distances versus time
(measured in units of $P_0$) for run~5 in table~\ref{tab1}.  In this
case $N=40$ and $R_*/R_{max}= 9.396\times 10^{-5}$.  In that case also
there are only two planets bound to the star after about $8,000 P_0$.
The others have either been ejected or have collided with the central
star.  Our treatment of tides is too crude to be certain
about the orbital evolution
 of planets on very eccentric orbits
which have  grazing approaches to the star.  In our simulations,
such  approaches  result in the planet being lost.  As far as the
subsequent evolution of the system is concerned however, it does not
matter whether the planet indeed hits the star of gets circularized on
a close orbit.  In this run, evolution occurs on a shorter timescale
due to the larger number of planets.  Here the innermost planet has
$e \sim 0.9$ and $a \sim 0.03 R_{max}$, which is 3~au if $R_{max}=100$~au.

\begin{figure}
\centerline{
\epsfig{file=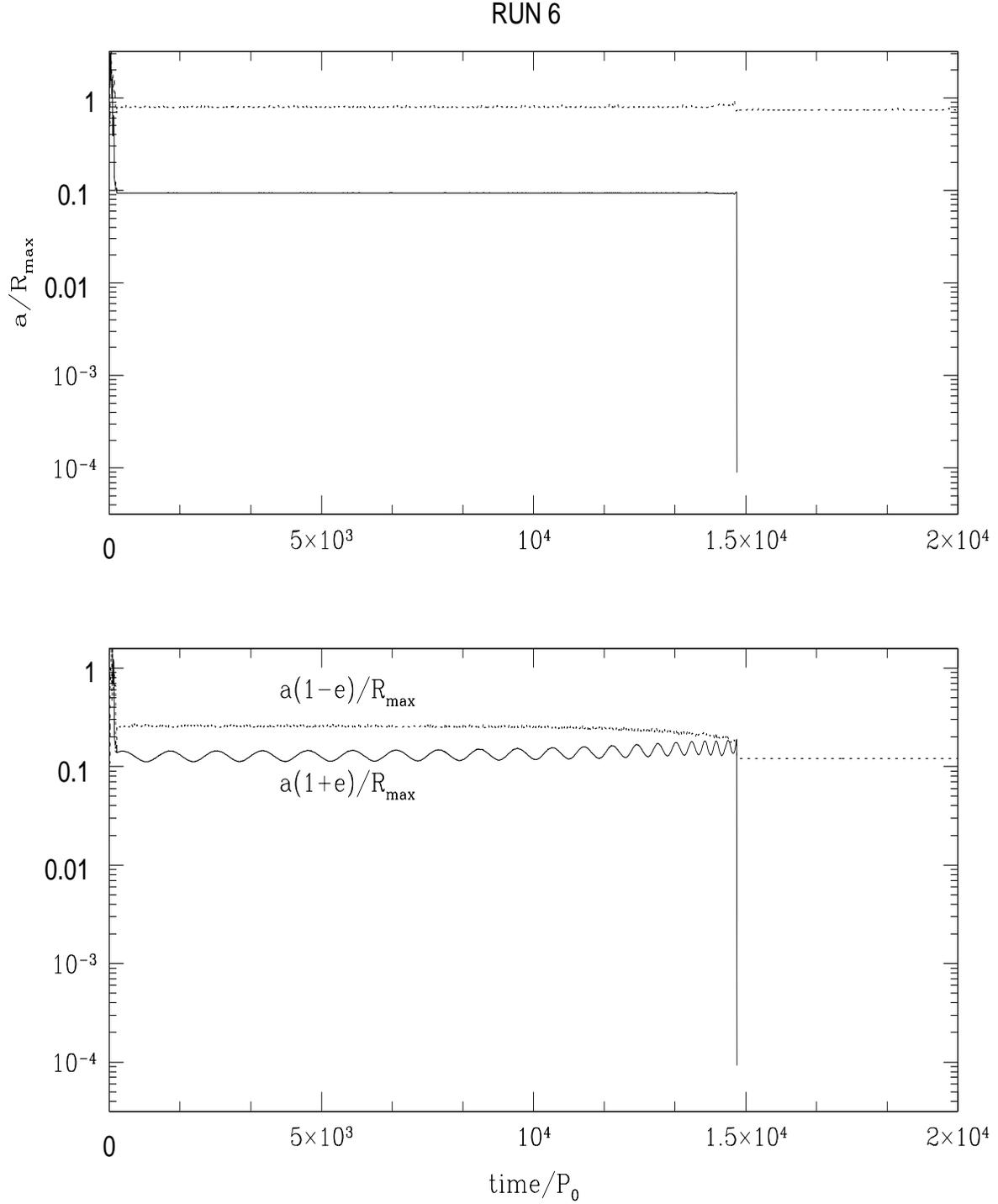,height=20.cm,width=16.cm} }
\caption[]{The {\em upper plot} shows the evolution of the semi-major
axes of the only 2 planets in the system which have not either been
ejected or collided with the central star after about 180~$P_0$ versus
time (measured in units of $P_0$) for run~6 in table~\ref{tab1}.  The
{\em lower plot} shows the pericenter distance ({\em dotted line}) and
the apocenter distance ({\em solid line}) for these planets.  Here
$N=100$, $R_*/R_{max}= 9.396\times 10^{-5}$ and the masses were
selected uniformly at random in the interval $(0;10^{-2} M_*).$ The
innermost planet suffers a gravitational scattering by the more
massive outermost planet when their orbits cross which results in an
increase of its eccentricity and eventually a collision with the
central star.  }
\label{fig6}
\end{figure}

Figure~\ref{fig6} shows the evolution of the semi--major axes versus
time (measured in units of $P_0$) for the only 2 planets in the system
which have not either been ejected or collided with the central star
after about 180~$P_0$, for run~6 in table~\ref{tab1}.  Also shown are
the pericenter distance and the apocenter distance for these planets.
Here $N=100$, $R_*/R_{max}= 9.396\times 10^{-5}$ and the masses were
selected uniformly at random in the interval $(0;10^{-2} M_*).$ We see
from Figure~\ref{fig6} that at a time of about $1.5 \times 10^4 P_0$ the
apocenter and the pericenter of the innermost and outermost planets,
respectively, are very close to each other.  The innermost planet then
suffers a gravitational scattering by the more massive outermost
planet which results in an increase of its eccentricity and eventually
a collision with the central star.  We are then left with one planet
with $e=0.85$ and $a \sim 0.7 R_{max}$.

\section{Summary and Discussion} \label{disc}

We have considered the orbital evolution of $N$ bodies with masses in
the giant planet range, which are assumed to have formed rapidly
enough that they can undergo subsequent dynamical relaxation on a
timescale $\sim 100$ orbits.

We have considered $5 \le N \le 100.$  We assume that rapid
formation may occur as a result of fragmentation triggered by
gravitational instabilities or clumping, either in a spherical envelope during
the initial protostellar collapse phase or in a disc like configuration
(e.g. Pudritz et al 1996) as that forms.  But note that as magnetic fields
may play a role the  disc may not be entirely
centrifugally supported. 
We have
considered  initial distributions of planets
with masses in the range 5 to 10 Jupiter masses that are in the form of a sphericalshell or are disc like.
However, final outcomes are independent of this.
In our calculations, along with  previous
work related to the
core accretion model (Weidenschilling \& Mazari 1996,  Rasio \& Ford  1996,
 Lin \& Ida 1997)
 we neglect the effects of gas,
thus  the efficiency of fragmentation is assumed to be  maximal.

Although our results
can be applied to different radial scales, we  focus the
discussion below on  distributions with an outer
radius of  100~au this is 
the dimension of observed protostellar discs in class 0 
objects (eg. Pudritz et al 1996).

This work has been partly motivated by the 
suggestion  by Black (1997) and Stepinski~\& Black (2000) that
some massive extrasolar planets could actually belong to
the low--mass tail of the
distribution of low--mass companions to solar--like stars.
That suggestion is derived from the observation that
extrasolar planets far enough from the star that tidal circularization
does not operate tend to have highly eccentric orbits
 and that these could be produced by a relaxation process of the type
we consider.

Altogether we have run about 25 cases with $30 \le N \le 100$ and many
more with $N \le 10$.  We have described six representative runs in
detail in the paper.  Independently of how many planets we began with,
the initial set up and the  technical details of the 
computation, we  obtained the same  sets of
characteristic behaviour and end states.
In every  case most of the planets where ejected
from the system and at most 3 planets remained bound to the central
star after a time typically on the
order of a few $10^4$ outer periods $P_0$.  The dynamical relaxation
phase is shorter when the initial number of planets is larger, but
when the number of planets  is reduced to less than 10 the system evolves
in the same way as the systems which had a smaller intital number of
planets.

We found that close encounters with the central star often occured
(for about 10\% of the planets) for all values of $N$ considered.  
At an early stage these tended to result in direct collisions.
When a direct collision is avoided, tidal interaction
between the star and planet on a very ecentric orbit may result in orbital
circularization at fixed pericentre distance which might ultimately
lead to the formation of a very closely orbiting giant planet.

As far as we could monitor the runs presented here, we did not find
any physical collisions between the planets themselves.  This is
consistent with Lin~\& Ida (1997) who found that such collisions were
very rare when, as here,  mutual tidal interactions between the planets were
assumed to be ineffective.  If physical collisions were to
occur , they would not be expected to affect the typical outcome
of our runs.

Typically the runs ended up with either (i) one potential 'hot Jupiter' plus 
 up to two 'external' companions, i.e.  planets orbiting near the edge
of the initial distribution; (ii) one or two 'external' planets or even
none at all since a gravitational scattering between two planets may
result in one being ejected and the other one colliding with the star;
(iii) one planet on an orbit with a semi--major axis 10 to a 100 times
smaller than the inital  distribution, e.g. $a \sim 1$--10~au for
a 100~au distribution, plus one 'external' companion.  Apart from the potential
'hot Jupiters', all these objects are on highly eccentric orbits.

We found that, apart for the 'hot Jupiters', the probability of ending
up with a planet orbiting at a given distance from the central star
increases with the distance.  Thus, this scenario produces a
decreasing number of planets as we go from 100~au down to 0.1~au.
This is expected since a planet on a highly eccentric orbit has more
chance of colliding with the central star as it gets closer to it.
Paradoxically the action of  the  central star tends to clear out
a large cavity around it. 

We now turn to the characteristics of some of  the extrasolar planets detected
so far.  We restrict our attention to those with a projected mass $M
\sin i > 4.5$~M$_J$, consistent with the larger
masses expected for formation through
fragmentation. 
There are 7 such isolated  objects around
HD~190228, UA3, HD~2222582, HD~10697, 70~Vir, HD~89744 and
HD~114762.  All have an eccentricity larger than 0.12, and for 6 of
them  $e \ge 0.33$.  The semi--major axis is in the range
0.88--2.5 for 5 of the objects, while  the others have $a=0.3$ and
0.43, respectively.  The characteristics of these planets  might
 be accounted for by our scenario
though the two cases with small $a$  would  only occur  with a small probability.

Amongst the 'hot Jupiters' detected so far, $\tau$ Boo
is particularly massive with $M \sin i \sim 4.$  
This planet is clearly a candidate
for being produced by a mechanism of the type we consider.

The  objects expelled as a result of the type
of relaxation process we consider
may produce a population of freely floating
planets ( Lucas \&  Roche 2000, Zapatero Osorio et al. 2000) which is
several times larger than that of giant planets close to the central
star.  This population would be expected to be  typically at
least 10 times larger than the population of massive planets orbiting
around the star, and depends on the initial number of planets in the
distribution.
However, note that the population of planets orbiting central stars
that went through a relaxation process of the type we consider
may be significantly smaller than the total if planets are also  independently
formed through a core accretion process.

The model we have considered in this paper has of course
some limitations.  In particular, we have not included the effect of
gas which is probably still significant in the early stages of
evolution of these systems, and which  may result in more planets
orbiting at smaller distances from the central star.  Also  we have
not considered the effects of Roche lobe overflow for the planets which
have a close encounter with the central star (Trilling et al. 1998)
which may significantly reduce their mass.

 The considerations above lead to the suggestion  that there may be
two populations amongst the extrasolar planets detected so far.  Some of
the more massive objects may
be produced through fragmentation of an
envelope or a disc--like structure followed by dynamical relaxation.
 Other  predominantly lower mass objects
could be produced in a disc as a result of the 
'core accretion' model.  We would of course expect to also have hybrid
systems, in which both processes have occured.  The relative
importance of these processes may depend for instance on  the
physical parameters, such as the angular
momentum content, of the parent cloud.

If the scenario presented here is indeed effective in planet formation,
we would expect  additional massive planets to be detected further from the
central star, and an important distribution of loosely bound, or
'free-floating' objects  associated with these systems.

\section*{acknowledgment}

J.C.B.P. acknowledges visitor support from the University Paris~6 and
the IAP where part of this work was carried out.  The authors thank
the participants of the EARA workshop 'Disks, Extrasolar Planets and
Brown Dwarfs' held at the IAP in July 2000 for useful discussions.

\onecolumn

\newpage

\end{document}